\documentclass[conferemce]{IEEEtran}
\usepackage{graphicx}
\usepackage{amsthm}
\usepackage{epsfig}
\usepackage{latexsym}
\usepackage{arydshln}
\usepackage{amsfonts}
\usepackage{here}
\usepackage{rawfonts}
\usepackage[latin1]{inputenc}
\usepackage[T1]{fontenc}
\usepackage{calc}
\usepackage{capitalgreekitalic}
\usepackage{url}
\usepackage{multirow}
\usepackage{enumerate}
\usepackage{color}
\usepackage[tbtags]{amsmath}
\usepackage{amssymb}
\usepackage{upref}
\usepackage{epic,eepic}
\usepackage{times}
\usepackage{dsfont}
\usepackage{comment}
\usepackage{cite}
\usepackage{bbm} 
\usepackage{multirow}
\usepackage{tikz}

\usepackage[pdfencoding=auto]{hyperref}
\hypersetup{
    colorlinks=true,
    linkcolor=blue,
    filecolor=blue, 
    urlcolor=black,
    citecolor=blue,
}

\usepackage{threeparttable}

\usepackage{dsfont}
\newcounter{step}
\newlength{\totlinewidth}
  {\end{list}%
  \rule{\linewidth}{1pt}}
\newcounter{substep}

  {\end{list}}

\newlength{\aligntop}
\newlength{\alignbot}
 

\IEEEoverridecommandlockouts

\usepackage{algorithm}%,algpseudocode}
\usepackage{algorithmic}

\makeatletter
\newcommand\semihuge{\@setfontsize\semihuge{19.3}{25}}
\makeatother

\makeatletter
\newcommand\semismall{\@setfontsize\semihuge{12.4}{15}}
\makeatother

\usepackage{subfigure}
\usepackage{tablefootnote}

\begin{document}

\title{\Huge{\color{black} Blockchains for Internet of Things: Fundamentals, Applications, and Challenges }\vspace*{-0em}}

\author{Yusen Wu, Ye Hu, {Mingzhe Chen}, Yelena Yesha, M\'erouane Debbah \vspace*{-2em}\\ 

\thanks{Y. Wu, M. Chen, Y. Hu, and Y, Yesha are with the University of Miami, FL, USA. M. Debbah is with KU 6G Research Center, Khalifa University of Science and Technology, P O Box 127788, Abu Dhabi, UAE and also with CentraleSupelec, University Paris-Saclay, 91192 Gif-sur-Yvette, France. This work was supported by the U.S. National Science Foundation under Grants CNS-2312139 and CNS-2312138.} %Email: \protect{mingzhec@princeton.edu}.}
 
 }

\maketitle
%
%%
%\vspace{0cm}

\begin{abstract}
\color{black}Internet of Things (IoT) services necessitate the storage, transmission, and analysis of diverse data for inference, autonomy, and control. Blockchains, with their inherent properties of decentralization and security, offer efficient database solutions for these devices through consensus-based data sharing. However, it's essential to recognize that not every blockchain system is suitable for specific IoT applications, and some might be more beneficial when excluded with privacy concerns. For example, public blockchains are not suitable for storing sensitive data. This paper presents a detailed review of three distinct blockchains tailored for enhancing IoT applications. We initially delve into the foundational aspects of three blockchain systems, highlighting their strengths, limitations, and implementation needs. Additionally, we discuss the security issues in different blockchains. Subsequently, we explore the blockchain's application in three pivotal IoT areas: edge AI, communications, and healthcare. We underscore potential challenges and the future directions for integrating different blockchains in IoT. Ultimately, this paper aims to offer a comprehensive perspective on the synergies between blockchains and the IoT ecosystem, highlighting the opportunities and complexities involved.
\end{abstract}

\section{Introduction}
 \color{black} As one of the most notable technologies in recent years, the Internet of Things (IoT) is expected to transform our daily lives with numerous applications such as smart homes, autonomous driving, smart grids, and more. However, the rise of IoT will also lead to a data explosion. Through a consensus-based data-sharing mechanism, blockchain techniques can provide IoT with a distributed, immutable, and efficient database solution, paving the way for a trustworthy data analysis and sharing platform.

Some existing works \cite{novo2018blockchain, wu2022bring, ray2020blockchain, satija2020blockene} have explored the use of blockchain in IoT applications. One of the most significant applications is edge artificial intelligence (edge AI) due to the increasing volume of data generated by edge IoT devices such as sensors and wearable devices. Processing data at edge devices can enhance training speed and data privacy, eliminating the need to exchange sensitive information, such as medical data, with other devices and central controllers, thus reducing the risk of data leakage. However, since edge devices are susceptible to various malicious attacks, utilizing blockchains to address data and model security concerns becomes essential. The application of blockchains in healthcare has also garnered considerable attention. A blockchain ledger's immutable and trusted nature ensures that transactions are secure, and its records are resistant to attacks. Additionally, permissioned blockchains can bolster data privacy for sensitive patient information. They can also be used to establish an electronic patient clinical trial system across different untrusted organizations, reducing traditional time-consuming administrative processes in hospitals. Lastly, leveraging blockchain for IoT communication stands as a crucial application in enhancing the security of data transmission across networks and IoT devices.

\color{black}In this article, we begin by introducing three fundamental blockchain systems: a) Public blockchain, b) Private blockchain, and c) Permissioned/consortium blockchain. We summarize their respective advantages, disadvantages, and implementation requirements. We also summarize the security issues associated with different blockchains. Next, we explore the application of these blockchain systems in three critical IoT domains: edge AI, healthcare, and IoT communications. Finally, we describe the challenges associated with implementing blockchain for IoT applications, concluding our discussion at the end.

{\color{black} Currently, several survey and magazine papers \cite{ dai2019blockchain, dedeoglu2019trust, hassan2019privacy,
 gadekallu2021blockchain, wang2019survey, fernandez2018review, wang2022integrating} have introduced the use of blockchains for IoT systems. However, this article is different from these existing papers in the following aspects. Compared to \cite{dai2019blockchain} and \cite{dedeoglu2019trust}, our paper details three distinct blockchains for different IoT scenarios, especially for permissioned blockchains. Different from \cite{hassan2019privacy} that ignored edge AI, we introduced permissioned blockchain for edge AI. Compared to \cite{gadekallu2021blockchain,wang2019survey,fernandez2018review,wang2022integrating} which discovered vast opportunities in the IoT domain, our paper points to several differences: 1) We introduce several unique blockchain applications for the Internet of Medical Things, for example, trusted executed environment (TEE) based private smart contract for clinical trials. 2) We have introduced choosing the appropriate blockchains for IoT applications, which is not discussed in the current works. 3) We delved into real IoT applications and summarized more important challenges for practical IoT settings.}

\begin{table*}
\begin{center}
\caption{Different types of Blockchain systems}
\begin{tabular}{ |c|c|c|c|}
 \hline
 Property & Public blockchain &Permissioned / Consortium blockchain & Private blockchain\\ \hline \hline
 Consensus determination   & All miners    &Invited and authorized members&  Administrator  \\ \hdashline[1pt/1pt]
 Permission &  Public  & Could be public or restricted   &Could be public or restricted\\ \hdashline[1pt/1pt]
 {\color{black}{Scalability}} &  {\color{black}{Typically low to moderate}}  & {\color{black}{Moderate to high}}   &{\color{black}{High}}\\ \hdashline[1pt/1pt]
 {\color{black}{Security}} &  {\color{black}{High}}  & {\color{black}{Moderate to high}}   &{\color{black}{Low}}\\ \hdashline[1pt/1pt]
 Immutability & Nearly impossible to tamper & Could be tampered with fewer nodes&  Could be tampered with fewer nodes \\ \hdashline[1pt/1pt]
 Efficiency {\color{black}{(latency,cost,throughput)}}    & Poor (high latency) & {\color{black}{Moderate to Good}} &  Good (low latency) \\ \hdashline[1pt/1pt]
 Centralized&   No  & Partial& Yes (administrator) \\ \hdashline[1pt/1pt]
 Consensus protocol & Proof of "X" (PoX)   & Byzantine fault tolerance (BFT)   & Byzantine fault tolerance \\ \hdashline[1pt/1pt]
 Membership & Dynamic  & Fixed; Know IDs of each other & Fixed; know IDs of each other \\ \hdashline[1pt/1pt]
 \multirow{3}{*}{\color{black} Applicability for IoT systems} & $\bullet$ {\color{black} High data security}  & $\bullet$ {\color{black} Supply chain management} & $\bullet$ {\color{black}High sensitive data privacy } \\ 
 &$\bullet$ {\color{black}Need transparency and trust } &$\bullet$ {\color{black} Healthcare and smart city}& $\bullet$ {\color{black}Require high transaction throughput }\\
 &$\bullet$ {\color{black}Need global reach by all users}& $\bullet$  {\color{black}Permissioned network}& $\bullet$ {\color{black}Low-cost network fees }\\
 &$\bullet$ {\color{black} Tokenization and micropayments}& $\bullet$ {\color{black}Require good transaction throughput}& $\bullet$ {\color{black}Private network participants} \\
 \hline
\end{tabular}
\label{tab:table1}
\vspace{-6mm}
\end{center}
\end{table*}

\section{Preliminaries and Overview}\label{se:system}

A  blockchain system simply is a shared and decentralized ledger that is publicly accessible by all the participants. A public immutable ledger is used to record transactions across many nodes (clusters). And the underlying consensus protocol ensures a new block added to the chains is in a final agreement. There are at least four types of blockchain networks in the industry: a) public blockchain, b) private blockchain, c) permissioned/consortium blockchain, and d) hybrid blockchain.  In this section, we mainly focus on the first three blockchains and explain their advantages, drawbacks, and implementation requirements as follows. Hybrid blockchain usually just combines and merges them in different requirements. A detailed comparison is shown in Table \ref{tab:table1}.

\subsubsection{Public Blockchains}
In a public blockchain system, a participant (whether a miner or a user) can join the blockchain network from anywhere and at any time. A core consensus protocol is employed to ensure agreement, trust, availability, and safety. For instance, Bitcoin relies on proof-of-work (PoW), whereas Ethereum utilizes proof-of-stake (PoS). Users can employ a digital blockchain wallet to manage cryptocurrencies like Bitcoin or Ether {\color{black}(ETH)} for the purpose of exchanging funds and assets. {\color{black} Public blockchains are well-suited for applications requiring both high data security and a necessity for data transparency.}

{\color{black}While the transparency and immutability of public blockchains make them resilient to censorship and tampering, we continue to witness incidents where attackers successfully steal cryptocurrencies as public blockchains are under severe attacks, such as transaction verification vulnerabilities (e.g., double spending, 51\% attacks, privacy, etc.), private key vulnerabilities, phishing attacks, wallet vulnerabilities, exchange vulnerabilities, physical theft, etc. To enhance security, we can utilize reputable wallet software, store private keys securely offline, enable two-factor authentication wherever possible, avoid sharing private keys or sensitive information, and keep your devices and software updated with the latest security patches.} {\color{black}{Moreover, efficiency includes the speed of processing transactions (latency), the energy and computational resources required (cost efficiency), and the system's overall ability to handle large volumes of transactions effectively (throughput). However, due to the PoW consensus, efficiency of the public blockchains are poor (high latency and low throughput).}}

\subsubsection{Private Blockchains} 
A private blockchain is overseen by a network administrator, and participants require consent or authorization from the administrator to join the network. Due to the fewer nodes in a private blockchain compared to a public blockchain, a private blockchain expends less time on achieving a final consensus {\color{black}(high efficiency)}. Consequently, it conserves considerable resources and provides significantly lower transaction fees with highly efficient transaction execution times. It finds applications in secure data exchange and the protection of highly sensitive medical data among small groups. However, because a private blockchain is a centralized network with a limited number of nodes, the data and transactions recorded in the ledger are comparatively less secure when contrasted with a public blockchain.

{\color{black}Security in private blockchains relies heavily on \textit{access control} and \textit{encryption}. Ensuring that only authorized participants can access the network and that data is encrypted at rest and in transit is crucial. Additionally, centralized control requires trust in the governing entity, however, it introduces risks if the governing body is not trustworthy. Establishing clear governance rules and mechanisms or implementing transparency and accountability in decision-making processes can be a solution to mitigate centralization risks. Moreover, even in private blockchains, smart contracts can have vulnerabilities that could be exploited by malicious participants (e.g., smart contracts are transparent to every participant). Therefore, conducting thorough security audits of smart contracts before deployment or implementing a monitoring tool to detect smart contract functions can be used to address emerging vulnerabilities.}

\begin{figure*}
\centering
  \includegraphics[width=0.7\linewidth]{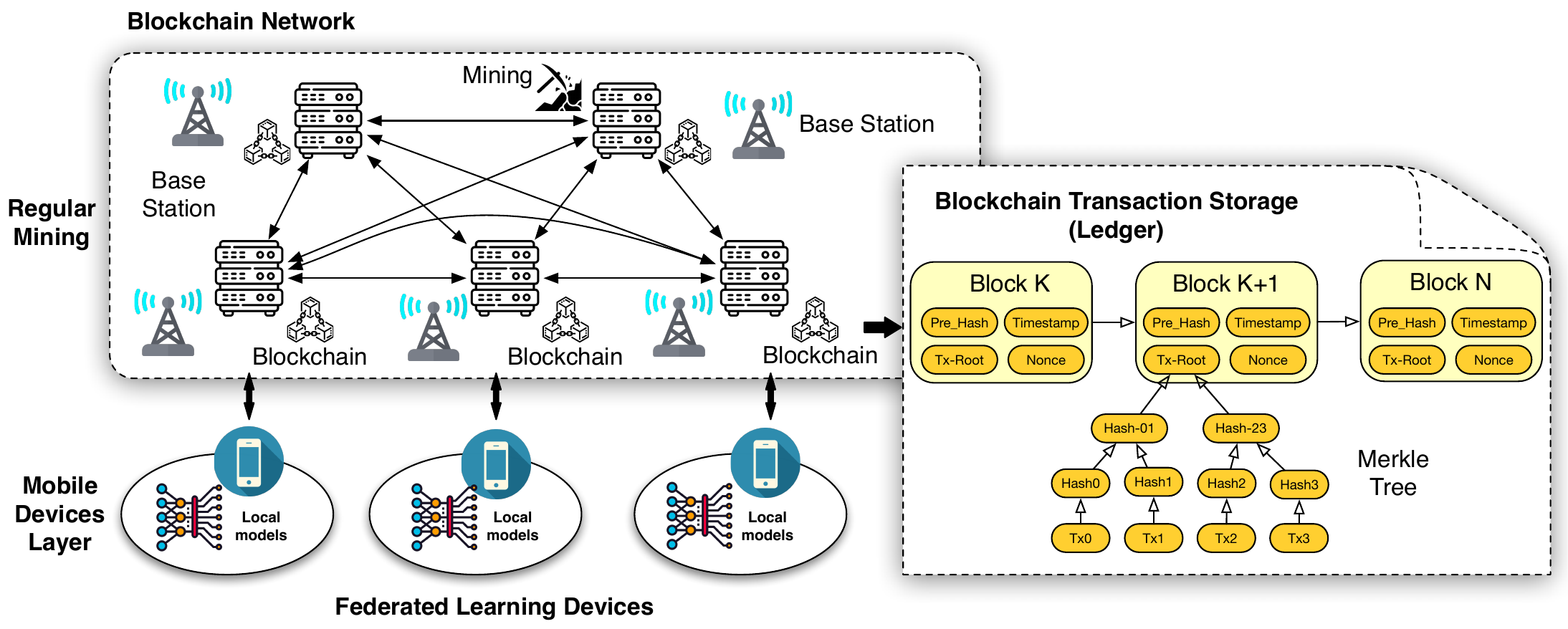}
  \caption{Blockchain for federated learning.}
  \label{fig:flearning}
\end{figure*}

\subsubsection{Permissioned / Consortium Blockchains} 
A permissioned blockchain, also referred to as a consortium blockchain, operates as a distributed ledger that is not openly accessible to the public. To become a participant in such a blockchain, one must be invited and gain acceptance from the majority of existing participants. In permissioned blockchains like Hyperledger Fabric or Iroha, each user's identity is both manageable and well-defined, and it is known to other participants. User access control is granted through an identity management module.

A permissioned blockchain represents a partially centralized network. When new members join the network, they also become owners of the blockchain and are included as part of the blockchain's committees. The underlying protocol of a permissioned blockchain typically employs a Byzantine fault tolerance (BFT) consensus protocol for achieving total order agreement {\color{black}{so that the efficiency is ensured}}. However, it's important to note that BFT consensus has limitations in terms of scalability, as latency tends to increase exponentially as more nodes join the network, and it requires at least three rounds of communication to reach a final agreement.

{\color{black}Security problems in permissioned blockchains involve assessing the robustness of the consensus mechanism and the composition of the validator nodes. In these networks, \textit{identity management} and \textit{authentication} are critical aspects of security, as ensuring the legitimacy of participants is fundamental to the system's integrity. Therefore, the implementation of robust access control mechanisms, including user authentication and role-based access control (RBAC) \cite{igarss}, is indispensable. Moreover, smart contract security, data privacy, network security, and compliance with regulatory standards are important considerations in permissioned blockchains, particularly in industries like finance and supply chain management. }

\section{Blockchains for Internet of Things}\label{se:metrics}
We introduce three main IoT applications of blockchain: 1) blockchain for edge AI, 2) blockchain for IoT communications, and 3) blockchain for IoT healthcare.

\subsubsection{Blockchain for Edge AI}
Edge AI enables distributed edge devices to collaboratively train a machine learning model without data sharing. However, since edge devices are vulnerable to malicious attacks such as {\color{black} poisoning and backdoor attacks}, the number of security issues and concerns in edge AI increases significantly compared to centralized AI. For example, within the edge AI training process, malicious edge devices may be compromised and send faulty models or gradients to a server or pretend they are correct among all the devices. Meanwhile, in edge AI, attackers among a large number of edge devices cannot be easily detected and traced since {\color{black} most of the attacks are anonymous}. To improve training security, privacy, traceability (explainable AI), and data integrity, {\color{black}permissioned blockchain seems a promising solution to public blockchain}. In particular, {\color{black} permissioned blockchain} can create a secure, decentralized, and highly sensitive system for storing and processing AI-generated data, models, or weights in the ledger so that all the data in the transactions become traceable. Meanwhile, {\color{black}permissioned blockchain} can use its analytics engine (failure detection) to block malicious identities thus improving model and data security. We introduce two specific applications of permissioned blockchain for edge AI as follows:

\begin{itemize}
    \item  \textit{\color{black}TEE-based smart contracts for privacy-critical and data-sensitive edge AI.} In permissioned blockchains, new participants must obtain permission before joining. This structure enhances data privacy in sensitive scenarios, such as maintaining the confidentiality of patient data across different organizations. {\color{black}Hospitals might employ a permissioned blockchain network for secure data exchange or to exclusively share AI models trained from edge devices, such as wearable health monitors that can incorporate ML to analyze health metrics in real-time, minimizing data leakage risks.} These models can be cataloged in a ledger using a unique model ID, model hash, and accompanying metadata. Through this ledger, hospitals can write, query, or aggregate different AI models, thereby establishing a private model-sharing platform fortified with smart contracts. Due to the immutable nature of the blockchain ledger, every AI training epoch can be chronologically documented. This not only ensures traceability in AI training but also guarantees that the parameters in the ledgers remain intact. Such immutability provides clarity on the derivation of the final AI model and flags potential malicious datasets. However, in a permissioned blockchain, smart contracts, although secure, are typically transparent to all members. This openness can pose a risk: inquisitive participants (though not necessarily malicious) could potentially access data via these contracts. To mitigate this, the use of private smart contracts is recommended. Given that smart contracts can be a vulnerable component in blockchain, safeguarding them with a trusted execution environment (TEE) ensures that data inputs stay concealed from even blockchain clients. In essence, a TEE-based smart contract enclave executes all its functions within a secure TEE zone, shielding all data even from the operating system and administrative users. Therefore, the synergy of edge AI and private smart contracts can notably elevate the levels of data privacy and security.
    \item \textit{Federated Learning meets permissioned blockchain in the edge.} Federated learning (FL) enables mobile devices to collaboratively learn a shared prediction model while keeping all the training data private. However, the devices are vulnerable to high-severity attacks, such as DDoS or backdoor attacks, which severely destroy the availability, security, correctness, and liveness of an FL framework. Federated learning can effectually take advantage of blockchain to bring trust to edge model, and improve training and model security, as shown in Fig \ref{fig:flearning}. Each device needs to register an identity in a blockchain network and thereby, each device owns a unique identity, and blockchain only accepts data from authorized devices.  {\color{black}Edge devices can collect data, analyze and train data, and then submit the gradients or parameters to the blockchain ledger as logs or receipts to make training traceable. In other words, leveraging blockchain enables edge models accessible and trusted}. Interestingly, one of the most significant challenges for data owners is that they always doubt how and when AI-based applications will use their data in federated learning, which means that there is no trust between data owners and systems. Leveraging blockchain technology, the data owners can fully control their data by giving permission or consent (e.g., electronic smart contracts including who and when can access the data) in AI-based applications.  Data owners can surface their data ownership by licensing their data to the application and storing all the policies or rules in immutable blockchain ledgers based on different conditions. {\color{black}Importantly, FL prioritizes data privacy since clients are often reluctant to reveal their datasets. As a result, public blockchains, which inherently prioritize transparency and openness, are not the ideal choice for FL.}
\end{itemize}

\subsubsection{Blockchain for Wireless Communications} \label{wire}
 
Motivated by the use of cryptocurrency and the establishment of secure decentralized networks among various untrusted entities, blockchain technology has been integrated into wireless networks \cite{chen2021distributed}. This integration permits uncertain individuals to engage with each other across different network areas. {\color{black}Specifically, blockchain can instill immutable trust among these distinct networks. For instance, public blockchains can ensure secure data access and transaction traceability. Conversely, permissioned or private blockchains can enhance data privacy and support private resource sharing in wireless networks.} We have listed several research areas related to the application of blockchains in wireless networks as follows:

\begin{figure*}
\centering
  \includegraphics[width=0.70\linewidth]{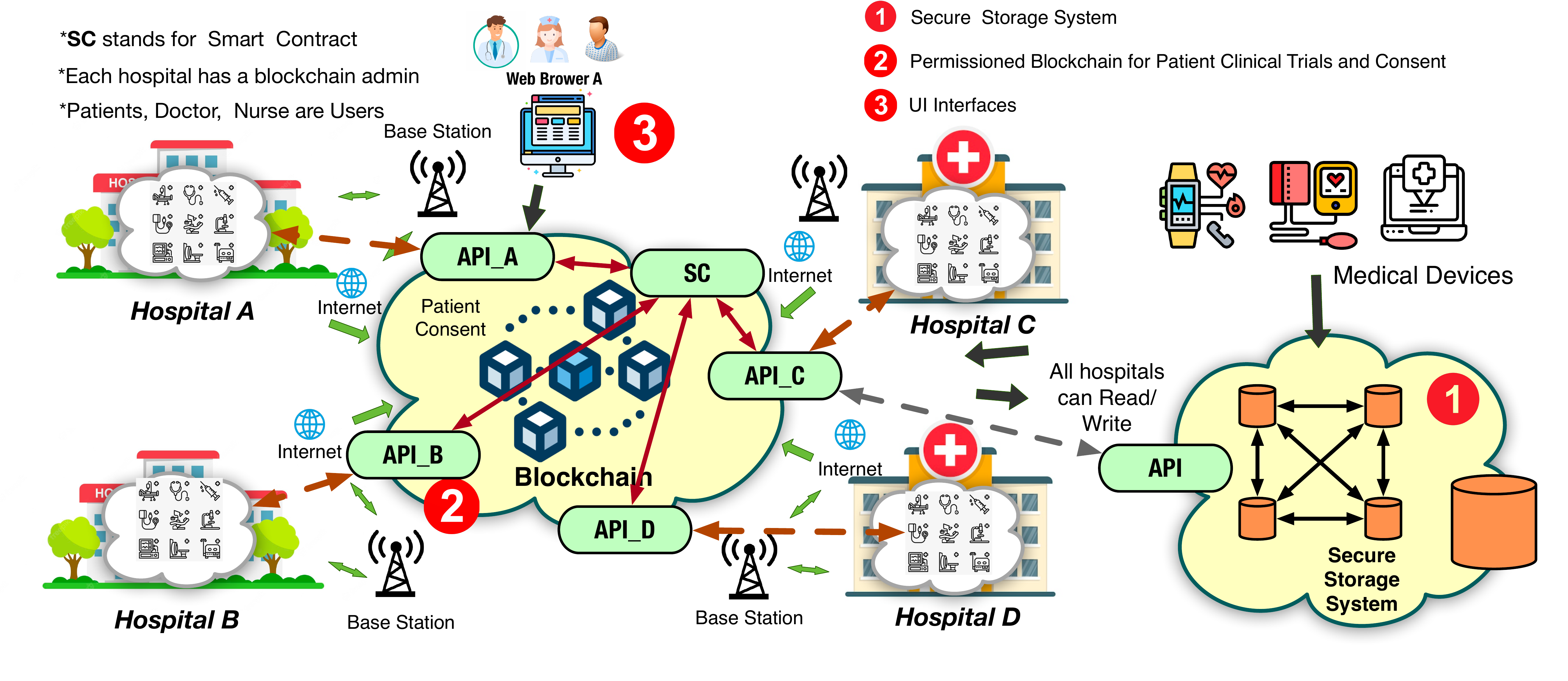}
  \caption{Healthcare platform for secure image exchange with permissioned blockchain.}
  \label{fig:hopsitals}
\end{figure*}

\begin{itemize}

   \item  \textit{Application management in wireless networks with blockchain.} Wireless devices include GPS, wireless computer devices (e.g., headphones or keyboards), satellites, cordless phones, radio transceivers, etc. The applications of wireless networks include different Wi-Fi applications or network services from network providers installed on those devices. A {\color{black}public} blockchain enables different providers to manage their services and assets efficiently, for example, by selling network services online through cryptocurrencies in an Ethereum blockchain, and the network will be active only after the provider gets a valid payment. Customers and providers can fully trust each other through the immutable and {\color{black}transparent} transactions in the ledgers. Likewise, a detailed and well-designed smart contract needs to be developed automatically to execute, control, or document legally relevant payments and events.
    
   \item  \textit{Adversarial modeling in wireless communications.} 
    Adversarial modeling is the technique of identifying malicious attackers based on some suspicious behaviors (e.g., time, entities, frequency, and protocols to thwart attacks), rather than only searching for specific indicators of an attack. Traditional solutions such as authentication and signature fail in this case since the attackers are often no anomalies and pretending they are honest, thereby adversarial modeling becomes the most common solution in wireless settings to filter spam and detect malware.  
    Combined with {\color{black}public} blockchain technology, the platform can act as an immutable notebook to record all the suspicious behaviors of attackers and become an adversarial detection toolbox for all users. A smart contract is designed and runs independently as a middleware to interact with blockchain ledgers. For example, invoking \textit{commit suspicious behaviors function} to store a behavior record including malicious behavior ID, timestamp, and descriptions, or \textit{query} a transaction by a transaction ID, or \textit{delete} a transaction.  
    
    Take blockchain in wireless vulnerability assessment as an example. Wireless security assessment helps identify vulnerabilities and security issues in the wireless settings, e.g., Wi-Fi vulnerability assessment can help detect and fix the risks before being attacked, we can discover nearby wireless devices, investigate rogue devices, test WLAN infrastructure, and apply the test results in a blockchain. A smart contract enables the monitoring of specific parts of the radio frequency (RF) spectrum to identify unauthorized wireless transmissions and activities. A dedicated smart contract also can be implemented to commit all the reports and results to immutable ledgers. Experts or researchers can join and contribute research results to the assessment. An intrusion detection report can also be part of the assessment. 
    
   \item  \textit{Secure routing protocols.} Secure routing protocols are designed to counteract routing attacks that might disrupt route discovery. A route fails when the source node uses up its network retries without receiving an ACK, and the routing poisoning attack generally happens by editing the routing tables. With {\color{black}public} blockchain, an ad hoc smart contract for managing and orchestrating \textit{iptables} can prevent them from malicious attacks, thus improving network routing security.

   \item  \textit{Digital forensics with blockchain in wireless communications.} Indisputably, {\color{black}public} blockchain technology can offer digital forensics with substantial benefits for the whole procedure, including identification, preservation, analysis, documentation, and presentation.  Wireless forensics similarly includes data collection, data analysis, security analysis, and network investigations to determine whether the network has been used for illegal purposes. The features of immutability, transparency, and trust of public blockchain bring reliable digital forensics to wire communications.
    
   \item  \textit{User privacy with {\color{black}permissioned} blockchain in wireless communications.} Blockchain transactions can be encrypted by data owners using private and public keys, enabling them to store the ciphertext privately in a ledger. Third-party participants, even authorized members within the same permissioned blockchain network, are prohibited from misusing or obtaining the data. When personal data is stored on the blockchain ledger, data owners have complete control over when and how a third party accesses it. This undoubtedly enhances user privacy and data confidentiality. Specifically, the advantage of employing cryptographic primitives to encrypt data during transmission from devices to servers via wireless networks is that it bolsters data privacy, especially when the raw data (which lacks privacy measures) is susceptible to eavesdropping attacks. {\color{black}Permissioned blockchain} participants can log all transactions in the ledger, efficiently manage or share their data, and utilize {\color{black}permissioned blockchain} technology to develop innovative data-sharing and access control applications.

\end{itemize}
    
\subsubsection{{\color{black} Permissioned} Blockchain with the Internet of Medical Things} 
Blockchain's adoption in healthcare is expanding rapidly with platforms like BurstIQ helping healthcare entities securely manage vast patient data. BurstIQ's platform offers comprehensive patient records, potentially aiding in curbing prescription drug misuse.  {\color{black} The Internet of Medical Things (IoMT) \footnote{The internet of medical things (IoMT) is the network of Internet-connected medical devices, hardware infrastructure, and software applications used to connect healthcare information technology} allows hospitals real-time patient connectivity, enabling timely data acquisition by healthcare professionals. Given the sensitivity of patient data, permissioned blockchains are preferred for ensuring data privacy \color{black}{due to the ability to ensure security, regulatory compliance, and efficient handling of sensitive patient data within an access controlled and customizable network}.}

\begin{figure}[t]
  \centering
  \includegraphics[width=1\linewidth]{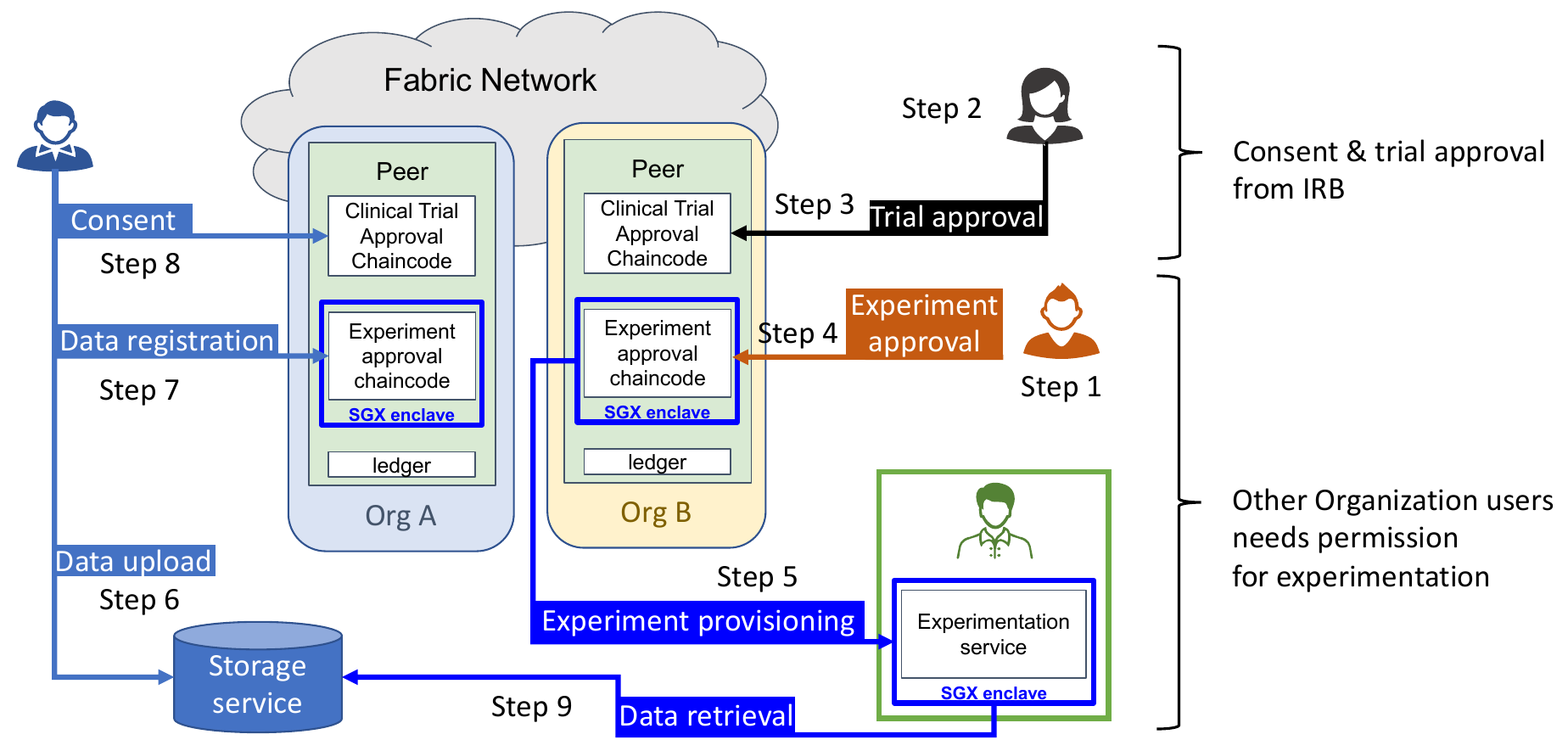}
  \caption{Secure patient clinical trials and consent system workflow overview with private smart contracts.}
  \label{fig:workflowoverc}
\end{figure}

\begin{itemize}
\item \textit{A zero-trust data sharing platform with IoMT.} Medical devices produce vast amounts of sensitive patient data. Therefore, a big data analytics platform is crucial for managing these valuable patient details, encompassing modules for data sharing, analysis, and storage. A demonstration of data sharing in an IoT-based healthcare system among various hospitals is depicted in Fig. \ref{fig:hopsitals}. Smart medical devices, such as robotic surgeons or wearable sensors, transmit diagnostic and therapeutic procedure data as transactions to the smart contract. This contract then records the transactions in the blockchain ledger. Various hospitals can participate as authorized members, enhancing specific medical research domains within the blockchain. Researchers or experimenters can access, download, and analyze the data once they obtain permission (a demonstration of how a third party obtains permission is discussed in section \ref{demo1}). Notably, since blockchains cannot store large CT scan images, such image data is uploaded to a secure or decentralized storage system \cite{duan2020intrusion} using API functions.
     
\item \textit{Secure lightweight solutions for IoMT systems with Blockchain.} 
Both public and permissioned blockchains require powerful servers for storage, consensus, and calculations. Lightweight solutions, such as a lightweight consensus protocol running on smart medical devices, aim to make all devices smarter by facilitating \textit{consensus overhead} between them. This is a research direction to reduce resources and costs because there might not be a need to set up many decentralized servers. In other words, a novel blockchain system designed specifically for IoT devices could be developed. Sambhav et al. \cite{satija2020blockene} proposed a new blockchain that runs a consensus module and data storage module separately (e.g., running consensus on IoT devices and storing data on regular servers), allowing smart devices to become a node in the blockchain system. While some lightweight cryptographic primitives can be used to simplify installation steps and ensure IoT data privacy, given that many devices have a limited configuration unsuitable for installing a full version of the cryptographic library, new consensus protocols still need to be developed for IoMT. This would account for limitations from medical devices (e.g., limited memory and storage) when thousands or millions of IoMT devices connect in the future.

\item \textit{Novel access control algorithms for blockchain-enabled IoMT.}
 The downsides of contemporary access control in IoT still suffer from complicated use and severe attacks. A blockchain and smart contract-based access control reduce cumbersome processes (e.g., how to choose a reasonable access control algorithm and deploy it). Permissioned blockchain, in particular, Hyperledger fabric blockchain, has its own attributed-based access rule for user identity management, and its \textit{channels} can efficiently set the access control policy for all the participants. 
     
\item \textit{{\color{black} Permissioned} blockchains with Electronic Health Record (EHR).} \label{demo1}
Centralized clinical trial systems, though commonly used, are often insecure and inefficient, especially when managing and sharing data across various disparate organizations. Ensuring patient and data privacy in such a centralized system presents challenges. Constructing a permissioned blockchain-based EHR application can effectively address concerns about data confidentiality, privacy, traceability, and integrity. We present a detailed patient clinical trial use case and its workflow to provide a clearer understanding of the innovative clinical trial architecture, as depicted in Fig \ref{fig:workflowoverc}.
\end{itemize}

\textbf{Workflow Overview.} {In this example, Institutional Review Boards (IRBs) approve clinical trials; investigators oversee them, while researchers monitor subjects and evaluate changes. HFB refers to the Hyperledger Fabric Permissioned Blockchain, and all its members (owner of blockchain nodes) are authorized participants.
Patients first register on the blockchain network. They grant consent using web interfaces (Step 8). Their sensitive data is stored and registered in an AWS cloud database or AWS S3 bucket (Step 6\&7), with only authorized users able to access it (Step 9). Specifically, third-party users need to send requests to ask for permission from patients. After obtaining patient consent (Step 1\&4), the smart contract provides a data endpoint link to the researchers (Step 5). Once they have the necessary approvals, they can download patient data for study. The complete implementation and evaluations are detailed in \cite{wu2023apply}.
Another example, when a clinician wants to conduct a clinical trial, he/she needs to submit a request to IRB, and IRB members approve a trial (Step 2\&3). }

\begin{comment}
  \begin{table}[t]
  \begin{center}
    \scriptsize
    \caption{Endpoints Response Time.}
    \label{tab:table2}
    \begin{tabular}{|l|c|c|} % <-- Alignments: 1st column left, 2nd middle and 3rd right, with vertical lines in between
      \hline
      \textbf{Request} & \textbf{Method} & \textbf{Latency} \\
      \hline \hline
      Register & POST & 233ms \\ 
      \hdashline[1pt/1pt]
      Consent Grant & POST& 5.23s\\
      \hdashline[1pt/1pt]
      Consent PatientId & GET & 623ms\\
      \hdashline[1pt/1pt]
      Consent Revoke & GET & 256ms \\
      \hdashline[1pt/1pt]
      Consent Acknowledge & POST & 516ms \\
      \hdashline[1pt/1pt]
      Consent & GET & 3.91s \\
      \hdashline[1pt/1pt]
      Consent Validate StudyNumber & GET & 4.64s\\
      \hdashline[1pt/1pt]
      IRB Trials (all trials) & GET & 4.97s  \\
      \hdashline[1pt/1pt]
      IRB Trials Register & POST & 4.36s\\
      \hdashline[1pt/1pt]
      IRB Trials Status & POST & 153ms\\
      \hdashline[1pt/1pt]
      IRB Trials Institutions & GET & 4.51s\\
      \hdashline[1pt/1pt]
      IRB Trials Join & POST & 751ms\\
      \hdashline[1pt/1pt]
      IRB Trials StudyNumber Status & POST & 45ms\\
      \hdashline[1pt/1pt]
      Hospital Trials StudyNumber Invitation & POST & 38ms\\
      \hline
    \end{tabular}
  \end{center}
\end{table}
  
\end{comment}

\section{Challenges}\label{se:communication}

While a public blockchain operates as an open ledger accessible to all, the advent of permissioned blockchains has addressed some significant challenges, particularly in terms of privacy and efficiency. However, permissioned blockchains still confront several issues. We delve into the primary challenges for both permissioned and public blockchains below.

%\subsection{Code development, installation, and management}
%Both private and permissioned blockchains need an administrator to first set up node and block which is relatively easy, but how to manage the whole system and make it reliable and bug-free needs more debugging and testing. Even deploying on cloud services, such as AWS blockchain, we still need time to learn the complicated platform. A Terraform \cite{shirinkin2017getting} automation tool can be used to set up the whole blockchain system in one click, but we need to learn Terraform languages and their platform usage. As the blockchain is in the early stage, a dedicated used blockchain needs well-designed smart contracts to manage and analyze transactions. A front end and API functions also need to be developed for different IoT applications. 

\subsection{Energy Efficiency}
When integrating public blockchains with IoT applications, a primary concern is energy efficiency. This concern spans beyond the energy demands of the IoT applications and data transmissions themselves and extends to the energy used in node competition stemming from consensus protocols. In many IoT systems, edge devices operate on limited energy resources, which can preclude them from serving as full consensus nodes or from handling intensive computational tasks. Additionally, certain public blockchain consensus protocols, like Proof of Work (PoW), require all participating miners (or devices) to constantly compete in solving complex mathematical puzzles, a process that markedly elevates energy consumption. {\color{black}Given these considerations, permissioned blockchains could play a pivotal role in enhancing data communication efficiency in IoT settings. }  

%As the PoW protocol needs all the miners to keep competing in solving math puzzles, energy efficiency becomes the most challenging problem for researchers. The US White House published estimates of the total global electricity usage for crypto-assets in 2022. The results are between 120 and 240 billion kilowatt-hours per year. Proof of stake (PoS) protocol requires much less energy, and no specialized equipment is needed. As a result, it is considered a more environmentally-friendly alternative to PoW, but it hasn't been proven to be as secure and stable as PoW at scale.

{\color{black}\subsection{Scalability in Blockchains}

{\color{black}Blockchains must be scalable to handle rising transactions and more network nodes. Scalability in blockchain technologies is significantly influenced by factors such as network size, consensus mechanism, block size and frequency, degree of decentralization, and the level of trust among participants. Public blockchains face scalability challenges due to the extensive validation processes required by a large number of nodes, using computationally intensive consensus mechanisms like PoW. In contrast, permissioned and private blockchains can achieve higher scalability through more efficient consensus protocols and smaller, more trusted networks that allow for quicker validation processes. Adjustments in block size and intervals can enhance throughput, but they must balance against potential security risks and the impact on decentralization. Thus, each blockchain type makes specific trade-offs to optimize scalability while considering security and operational needs.}

{\color{black}
\subsection{Pruning in Permissioned Blockchains}
The size of the blockchain continuously grows as new blocks are added. Over time, this can lead to significant storage requirements, which can be especially challenging for nodes that need to store the entire blockchain. The larger the blockchain, the longer it may take to validate, synchronize, and back up the data. This can affect the performance and speed of the entire network. In order to solve the problem, implementing \textit{garbage collection} mechanisms to identify and remove data that is no longer required. This is especially applicable in permissioned blockchains where certain data might have a defined "active period" or "active duration".}

{\color{black}
\subsection{Integration with Existing IoT Systems}
Integrating public or permissioned blockchains with current IoT infrastructure presents challenges. For instance, different blockchains have different protocols, consensus mechanisms, and data structures. Integrating multiple blockchains or connecting a blockchain with a legacy system might require complex bridging solutions. However, there are emerging solutions to address these challenges, including novel middleware for blockchain integration, Blockchain as a Service (BaaS) offerings, and standardized APIs for interacting with legacy systems. Crucially, ensuring that the entire system is reliable and free from defects requires extensive debugging and testing. Even when deploying on cloud platforms like AWS blockchain, there's a learning curve involved due to the platform's complexity.}

{\color{black}
\subsection{Interoperability for Private Blockchains}
How do private blockchains communicate and interact with other private blockchains or even public blockchains? The challenge of ensuring seamless communication between disparate blockchain systems is often referred to as "interoperability". Several solutions have been proposed, such as cross-chain communication protocols, atomic swaps between different blockchain systems, and standardized APIs. However, trust issues, security concerns, or data and protocol mismatches still are the most significant barriers.}

{\color{black}
\subsection{Incentivization Mechanisms in Permissioned Blockchains for Data Owners}
In public blockchains, network participants (like miners or validators) are incentivized through monetary rewards (like mining or staking rewards) to validate and add transactions to the blockchain. However, permissioned blockchains generally lack this direct financial incentive, as they are typically used by consortiums or groups of known entities for specific purposes and do not rely on native tokens or cryptocurrencies for operation. There are still ways to motivate participation, such as adding reputation. The key is to identify what's valuable for the participants and structure the incentives around that.}

\section{Conclusion}\label{se:conclusion}
In this paper, we presented a thorough review of various blockchain applications for IoT. Initially, we provided an overview of three distinct blockchain systems, outlining their strengths, weaknesses, and implementation needs. We also summarize the security issues associated with different blockchains. Subsequently, we delved into three primary blockchain applications: edge AI, IoT communications, and IoT healthcare, illustrating their function across different blockchain platforms. In conclusion, we discussed the challenges and future research directions for blockchain's integration within IoT contexts.

\section*{Acknowledgments}
This work was supported in part by the U.S. National Science Foundation under Grants CNS-2312139 and CNS-2332834.

\bibliographystyle{IEEEbib}
\def\baselinestretch{0.96}
\bibliography{references1}
\footnotesize
\vspace{0.5cm}

\vspace{0.2cm} 
 \noindent {\bf Yusen Wu} (\href{mailto:yxw1259@miami.edu}{yxw1259@miami.edu})  is an Assistant Scientist at the Institute for Data Science and Computing (IDSC), University of Miami, USA. His research interests include distributed systems, blockchain, and AI for healthcare.
 
 \vspace{0.2cm}
\noindent {\bf Ye Hu} (\href{mailto:yehu@miami.edu}{yehu@miami.edu}) is currently an Assistant Professor at the University of Miami. Her research interests include wireless networks, machine learning, game theory, unmanned aerial vehicles, and cyber-physical systems.
 
 \vspace{0.2cm}
\noindent {\bf Mingzhe Chen} (\href{mailto:mingzhe.chen@miami.edu}{mingzhe.chen@miami.edu})  is currently an Assistant Professor at the Electrical and Computer Engineering Department, University of Miami, USA. His research interests include machine learning, virtual reality, unmanned aerial vehicles, and wireless networks.

 \vspace{0.2cm}
\noindent {\bf Yelena Yesha} (\href{mailto:yxy806@miami.edu}{yxy806@miami.edu})  is the Knight Foundation Endowed Chair of Data Science and AI at the Institute for Data Science and Computing (IDSC), University of Miami, USA. Dr. Yesha is also the Innovation Officer and Head of International Relations. She is endowed professor both in Department Computer Science and Department of Radiology at UM. Her research interests include blockchain, big data analytics, and AI for healthcare.
 
 \vspace{0.2cm}
\noindent {\bf M\'erouane Debbah} (\href{mailto:email: merouane.debbah@ku.ac.ae}{merouane.debbah@ku.ac.ae})   is a full professor at KU 6G Research Center, Khalifa University of Science and Technology and CentraleSupelec, University Paris-Saclay, 91192 Gif-sur-Yvette, France. He has founded several public and industrial research centers, start-ups and is now Chief Researcher at the Technology Innovation Institute in Abu Dhabi. He is a frequent keynote speaker at international events in telecommunication and AI. His research has been lying at the interface of fundamental mathematics, algorithms, statistics, information and communication sciences, with a special focus on the applications of random matrix theory and learning algorithms to communication sciences. In the wireless Communication field, he has been at the heart of the development of small cells (4G), Massive MIMO (5G) and Large Intelligent Surfaces (6G) technologies, for which he received multiple distinctions.
\end{document}